\immediate\write18{cp `kpsewhich apacite.bst` .}
\documentclass[twocolumn,prl]{article}
\usepackage[utf8]{inputenc}
\DeclareUnicodeCharacter{2010}{-}
\usepackage[T1]{fontenc}
\usepackage[a4paper, left= 1.2cm, right= 1.2cm, top=1.5cm, textwidth=15.5cm]{geometry}
\usepackage{graphicx}
\usepackage{geometry}
\usepackage{amsmath}
\usepackage{amssymb}
\usepackage{amsfonts} 
\usepackage{caption}
\usepackage{subcaption}
\usepackage{authblk}
\usepackage{caption}
\usepackage{subcaption}
\usepackage[backend=biber,sorting=none,style=numeric-comp,giveninits=true,doi=false,isbn=false,url=false,eprint=false,maxbibnames=99,date=year]{biblatex}
\renewbibmacro{in:}{}
\DeclareFieldFormat{pages}{#1}
\DeclareSourcemap{
  \maps[datatype=bibtex]{
    \map{
      \step[fieldsource=issue, match=\regexp{\A[0-9]+\Z}, final]
      \step[fieldset=number, origfieldval]
      \step[fieldset=issue, null]
    }
  }
}
\renewbibmacro*{volume+number+eid}{%
  \printfield{volume}%
  \setunit*{\addnbspace}
  \setunit{\addcomma\space}%
  \printfield{eid}}
\DeclareFieldFormat[article]{number}{\mkbibparens{#1}}
\title{Glassy Dynamics in Polyalcohols: Intermolecular Simplicity vs. Intramolecular Complexity}
\author[a]{Till Böhmer}
\author[b]{Jan Philipp Gabriel}
\author[a]{Rolf Zeißler}
\author[a]{Timo Richter}
\author[a]{Thomas Blochowicz}
\affil[a]{Institute for Condensed Matter Physics, Technical University Darmstadt, Germany}
\affil[b]{Glass and Time, IMFUFA, Department of Science and Environment, Roskilde University, Denmark}
\date{}

\renewenvironment{abstract}{%
  \begin{center}%
    {\bfseries \vspace{-0.5em}\vspace{0pt}}%
  \end{center}%
  \quotation}
  {\endquotation}

\begin{document}
\twocolumn[
\maketitle
\vspace{-1.5cm}
\begin{abstract}
Using depolarized light scattering, we have recently shown that structural relaxation in a broad range of supercooled liquids follows, to good approximation, a generic line shape with high-frequency power law $\omega^{-1/2}$. We now continue this study by investigating a systematic series of polyalcohols (PAs), frequently used as model-systems in glass-science, i.a., because the width of their respective dielectric loss spectra varies strongly along the series. Our results reveal that the microscopic origin of the observed relaxation behavior varies significantly between different PAs: While short-chained PAs like glycerol rotate as more or less rigid entities and their light scattering spectra follow the generic shape, long-chained PAs like sorbitol display pronounced \emph{intra}molecular dynamic contributions on the time scale of structural relaxation, leading to systematic deviations from the generic shape. Based on these findings we discuss an important limitation for observing the generic shape in a supercooled liquid: The dynamics that is probed needs to reflect the \emph{inter}molecular dynamic heterogeneity, and must not be superimposed by effects of \emph{intra}molecular dynamic heterogeneity.
\end{abstract}
    \vspace{\baselineskip}
]

\section{Introduction}
Traditionally, polyalcohols (PAs) have been intensively studied as prototypical glass-forming liquids. Especially worth mentioning is the homologous series glycerol, threitol, xylitol and sorbitol, sharing the structure formula HO-CH$_2$-(CH-OH)$_n$-CH$_2$-OH with $n=1\ldots 4$. Reasons for their prototypical status are numerous: They are easily handled and supercooled, important properties vary significantly among the series, e.g., fragility or non-exponentiality of the $\alpha$-relaxation\,\supercite{Doess2002}, and they share certain chemical similarities with water, arguably the single most important liquid on earth. As a result, numerous important  glass-transition properties were investigated using PAs as model-systems. Namely, glassy cooperativity\,\supercite{Dixon1991,Wiltzius1991,Menon1995,Berthier2005,Crauste-Thibierge2010,Fragiadakis2011,Brun2012,Bauer2013,Uhl2019,Albert2021}, dynamic\,\supercite{Miller1997a,XiaoHuaQiu2003,Zondervan2007,Hong2011} and structural\,\supercite{Gabriel2021} heterogeneity, correlations of different properties with fragility\,\supercite{Ngai1998,Doess2002,Carpentier2003,Kubo2006,Hong2011a,Paluch2011,Ruta2012,Migliardo2017,Drozd-Rzoska2019,Chen2021}, characteristics of structural relaxation\,\supercite{Dixon1990,Bohmer1998,Doess2002,Ruggiero2017,Krynski2020,Sibik2014}, the nature of the $\beta$-process\,\supercite{Doess2002,Doess2002a,Hensel-Bielowka2002,Paluch2003,Power2003,Bohmer2006,Yardimci2006,Geirhos2018,Krynski2020,Sibik2014} and the excess wing\,\supercite{Schneider2000,Lunkenheimer2002,Gainaru2020,Ngai2020}, physical aging\,\supercite{Miller1997,Miller1997a,Leheny1998,MouraRamos2007,Roed2019} and properties of PAs as ultra-stable glasses\,\supercite{Capponi2012,Kasina2015}, are only a few of the most significant research aspects.

\begin{figure}
    \centering
    \includegraphics[width=0.45\textwidth]{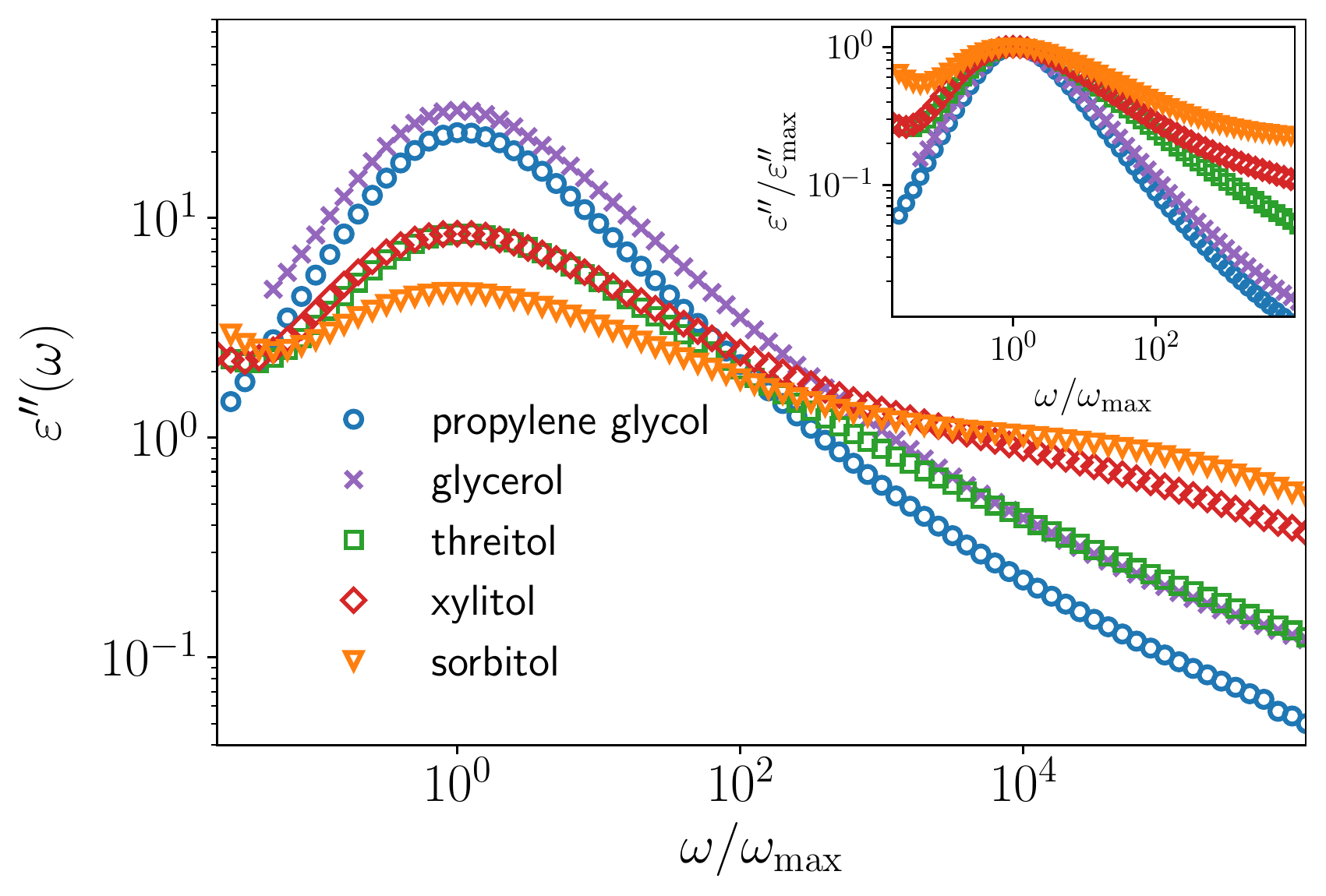}
    \caption{Dielectric loss data of different PAs normalized to their peak maximum frequency. Respective temperatures are 180\,K, 200\,K, 235\,K, 255\,K and 275\,K, from top (propylene glycol) to bottom (sorbitol). In the inset the same data are normalized to their peak maximum amplitude to compare their respective high-frequency power laws.}
    \label{fig:bds_shape}
\end{figure}

A majority of these experimental results is based on data from broadband dielectric spectroscopy (BDS), probably the most wide-spread technique to investigate the dynamics of supercooled liquids and molecular glasses. In such an experiment, the macroscopic dielectric response of the sample is studied as a function of the frequency of an applied electric field. The response is a function of the molecular dipole moments, thus allows access to the microscopic molecular dynamics.\,\supercite{Bottcher1978} In Fig.\,\ref{fig:bds_shape} we show dielectric loss spectra of different PAs, focusing on the $\alpha$-relaxation closely above the respective glass transition temperature. Strikingly, the shape of these spectra differs drastically between the different PAs, despite their chemical similarity. With increasing molecular weight, the width of the peak increases, while the slope $\beta$ of the high-frequency power law $\omega^{-\beta}$ decreases (cf. inset). Following the simple traditional interpretation, one would conclude from these two observations that dynamic heterogeneity increases with increasing molecular weight for PAs.

Based on findings from depolarized light-scattering (DLS), we recently challenged this simple approach of analyzing dielectric loss spectra. We showed that loss spectra of a broad range of different molecular glass-formers obtained in photon-correlation spectroscopy (PCS), in fact possess a common \textit{generic} shape with high-frequency behavior $\omega^{-1/2}$.\,\supercite{Pabst2021} In contrast, large variations are observed between the shapes of the respective dielectric loss spectra\,\supercite{Paluch2016,Pabst2021}. The latter was shown to result from the fact that slow dipolar cross-correlations can contribute significantly to the dielectric loss, whereas they mostly do not contribute to DLS. Those cross-correlations typically appear as a close-to single-exponential peak, superimposing the generic $\alpha$-process shape at lower frequencies and thus, lead to a broad range of observed spectral shapes for the dielectric loss.\,\supercite{Pabst2020,Pabst2021} The fact that loss spectra from DLS share a generic shape for many different glass-formers, suggests that the underlying distribution of relaxation times in supercooled liquids is also generic. This is surprising as these substances differ significantly regarding chemical structure and consequently, also regarding microscopic interactions.

In regard to this new perspective, PAs are interesting to study, due to the broad range of spectral shapes they display in BDS. Additionally, their prototypical status and their numerous experimental applications, call for an in-depth understanding of their properties. Therefore, the analysis of PAs using DLS will be the objective of the present study. So far, only glycerol has been intensively investigated in DLS\,\supercite{Gabriel2020,Wuttke1994,Lebon1997,Brodin2005}. In line with the aspects discussed above, the DLS spectra match the generic shape with a high-frequency power law of $\beta_\mathrm{DLS}=0.5$, whereas $0.57<\beta_\mathrm{BDS}<0.71$\,\supercite{Nielsen2009,Ngai2003} is found for the power law of dielectric loss data, depending on the analytical approach. From the comparison of BDS and DLS spectra, slow dipolar cross-correlations were identified in the BDS spectra.\,\supercite{Gabriel2020} Their origin is still unresolved, as both hydrogen bonding and dipole-dipole interactions could be relevant. In addition to glycerol, DLS spectra of propylene glycol (PG) have been shown to follow the generic shape\,\supercite{Pabst2021}, however until now, no detailed comparison with BDS data has been presented. For threitol, xylitol and sorbitol, to our knowledge, no DLS data have been published so far. The width of their dielectric loss spectra increases with increasing molecular weight and the high-frequency power law approaches values as low as $0.27<\beta_\mathrm{BDS}<0.30$\,\supercite{Nielsen2009} for sorbitol. It is therefore unlikely that the dielectric loss spectrum of sorbitol can be described by the generic shape and additional slow cross-correlations. However, there exist several peculiarities of molecular dynamics in higher molecular weight PAs that are possibly reflected in the respective dielectric loss spectra and make their interpretation non-trivial. To be specific, several effects that could be summarized as \textit{intramolecular dynamic heterogeneity} have recently been observed for sorbitol, which means different chemical bonds in the molecule rotate on different time-scales. In $^{13}$C-NMR and molecular dynamics simulations studies, dynamic decoupling was observed for different C-H bonds, with the bonds in the center of the carbon-backbone being slower than the ones at the ends\,\supercite{Margulies2000,Sixou2001}. A similar form of decoupling was observed between C-H and O-H bonds in $^2$H-NMR studies of -D$_4$ and -OD$_6$ labeled sorbitol (cf. Fig. 6.36 in Ref.\,\cite{Doess2001}). Detailed analysis of the differently labeled compounds revealed that not only their characteristic relaxation times, but also their respective spectral shape of the $\alpha$-process differs significantly in NMR\,\supercite{Becher2021}. These findings are especially relevant for interpreting BDS data, as the total molecular dipole results from the vector sum of all dipole moments of single C-O-H units in the molecule. Therefore, BDS does not probe one molecule-fixed axis in sorbitol, but rather an axis that depends on the current molecular conformation. In contrast, no dynamic decoupling was observed for glycerol, thus the dipole moment can be assumed to represent a fixed molecular axis in good approximation\,\supercite{Doess2001,Becher2021a}.

In this study, we want to address questions that are on the one hand crucial for the understanding of supercooled and glassy PAs but also relevant for the further development of the generic structural relaxation hypothesis:

(i) Do PAs of higher molecular weight follow the same generic shape as found for propylene glycol, glycerol and many other supercooled liquids?

(ii) Do slow cross-correlations contribute to the dielectric loss spectra of all PAs?

(iii) What are the limitations for observing the generic shape of the $\alpha$-process in a supercooled liquid?

To attempt an answer of these questions we perform DLS and BDS measurements for the discussed PAs and compare their respective spectral shapes.

\section{Experimental background}
Propylene glycol\,(PG, 1,2-propanediol, Alfa Aesar, 99.5\%), DL-threitol\,(Sigma Aldrich, 97\%), xylitol\,(Acros Organics, 99\%) and D-sorbitol\,(Acros Organics, 97\%) were measured in BDS and PCS. The data of glycerol were already published and discussed in Ref.\,\cite{Gabriel2020}. PG was used in this study, since ethylene glycol, the next smaller element of the homologous series after Glycerol, tends to crystallized and is not easily supercoolable. For the BDS measurements PG and glycerol were used as received. For PCS both substances were filtered using a 400\,nm hydrophilic syringe filter to remove dust, which is crucial for carrying out high-resolution PCS measurements.  The other PAs were received as a white powder. For BDS the powder was first filled into a sample cell, then dried at 1\,mbar and room temperature for one hour, and finally melted and once again dried at 150$^\circ$C and 1\,mbar for three hours. The PCS samples were prepared using the same drying and melting procedure, however carried out with the powder filled into a glass syringe. A stainless steel filter holder in combination with 450\,nm Nylon membrane filters was used to fill the liquid sample into the PCS sample cell at 150$^\circ$C. This procedure was repeated several times until a sufficiently large sample volume was obtained. Finally, the sample was again dried in the vacuum oven for more than three hours. 

For sorbitol, we also carried out high-temperature measurements, using a Tandem Fabry-Perot interferometer (TFPI). The respective sample was dried and melted in a 10\,mm glass tube using the procedure described above. Afterwards, the liquid sorbitol was quenched into the glassy state, using liquid nitrogen and finally, the tube was evacuated and flame-sealed to prevent oxidation at high temperatures. Using an Argon atmosphere instead of a vacuum yielded comparable results.

All BDS measurements were carried out using a Novocontrol Alpha-N High Resolution Dielectric Analyzer to probe the complex dielectric susceptibility $\varepsilon^*(\omega)$. PCS measurements were performed in vertical horizontal (VH) depolarized geometry at a 90$^\circ$ angle using a home-built setup described elsewhere in detail\,\supercite{Pabst2017,Gabriel2015,Blochowicz2013}. The intensity autocorrelation function of the scattered light $g_2(t)=\langle I_\mathrm{s}(t) I_\mathrm{s}(0)\rangle/\langle I_\mathrm{s}\rangle^2$ was obtained using a hardware correlator. From $g_2(t)$, the electric field autocorrelation function $g_1(t)=\langle E^*_\mathrm{s}(t) E_\mathrm{s}(0)\rangle/\langle |E_\mathrm{s}|\rangle^2$ was calculated using the Siegert relation for partial heterodyning, as described in Ref.\,\cite{Pabst2017}. Finally, PCS data were transformed into the susceptibility representation through Fourier transformation
\begin{equation}
    \chi^{\prime\prime}(\omega)\propto \int\limits^\infty_0g_1(t)\cos(\omega t)\mathrm{d}t,
\end{equation}
by using the Filon algorithm\,\supercite{Filon1929}. The TFPI data were obtained using a Sandercock TFPI analyzing backscattered light in VH geometry. From the measured spectral density $S(\omega)$, the susceptibility $\chi^{\prime\prime}(\omega)$ is calculated in a straightforward manner\,(cf. Ref\,\supercite{Gabriel2020}).

In the susceptibility representation, DLS data can be directly compared to the dielectric loss $\varepsilon^{\prime\prime}(\omega)$\,\supercite{Gabriel2018}. Both representations describe molecular reorientation, however one has to consider the different molecular properties that are probed. As discussed above, BDS probes the reorientation of the total molecular dipole moment, which is not necessarily fixed to a well-defined molecular axis, e.g. in PAs it can rotate significantly through comparatively small O-H bond motions (e.g. compare Table 3 in Ref.\,\cite{Alonso2018}). By contrast, DLS probes the reorientation of the optical anisotropy tensor, which is given by the anisotropic portion of the molecular polarizability \,\supercite{Berne1976}. All atoms of a molecule contribute to the latter, which therefore often is somewhat less affected by intramolecular motions as compared to the molecular dipole moment. 

\section{Results}
\begin{figure}
    \centering
    \includegraphics[width=0.45\textwidth]{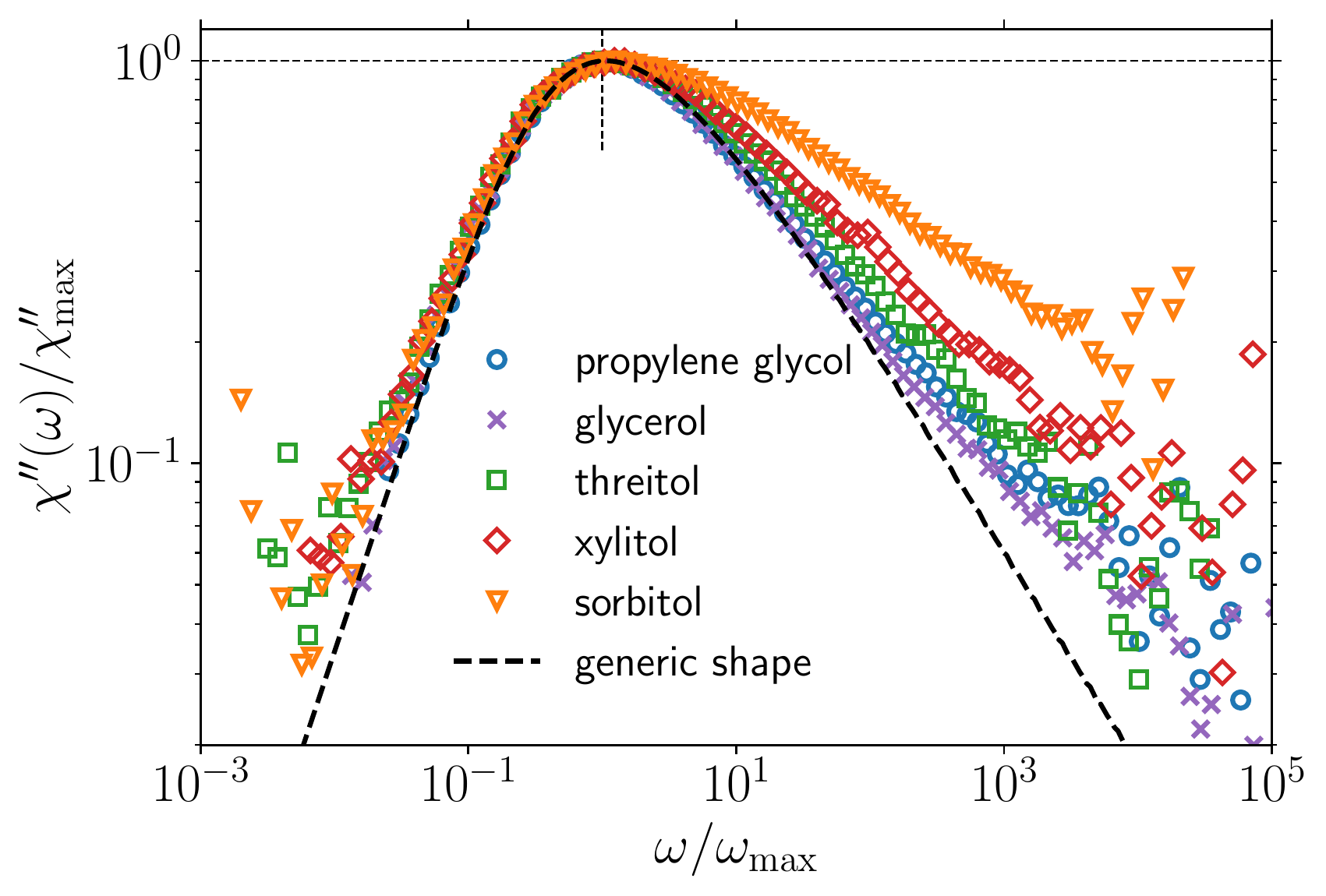}
    \caption{DLS data of different PAs in susceptibility representation, normalized to the peak maximum frequency and amplitude. Temperatures are 175\,K, 200\,K, 235\,K, 255\,K and 277.5\,K from top to bottom. The generic spectral shape from Ref.\,\cite{Pabst2021} is shown as the black dotted line.}
    \label{fig:pcs_shape}
\end{figure}
We begin by comparing DLS data of the different PAs in Fig.\,\ref{fig:pcs_shape}. For each PA, one representative susceptibility spectrum, close to $T_\mathrm{g}$, normalized to its respective peak maximum frequency and amplitude is shown. As already discussed in Ref.\,\cite{Pabst2021}, PG and glycerol follow a generic spectral shape (black dashed line) to good approximation up to frequencies where the typical excess wing develops. The latter feature was  shown to likely be related to an unresolved $\beta$-process\,\supercite{Schneider2000,Lunkenheimer2002,Gainaru2020}. Threitol, xylitol and sorbitol only follow the generic shape up to frequencies slightly above the peak maximum and deviate from it at higher frequencies. The deviation is systematic in the sense that it increases with increasing molecular weight, respectively number of OH-groups per molecule, of the PAs. While glycerol and PG follow a high-frequency power law with slope $\beta_\mathrm{DLS}=0.5$, the respective values roughly equal (0.40, 0.35 and 0.25)$\pm0.05$ for threitol, xylitol and sorbitol. 

\begin{figure}
    \centering
    \includegraphics[width=0.45\textwidth]{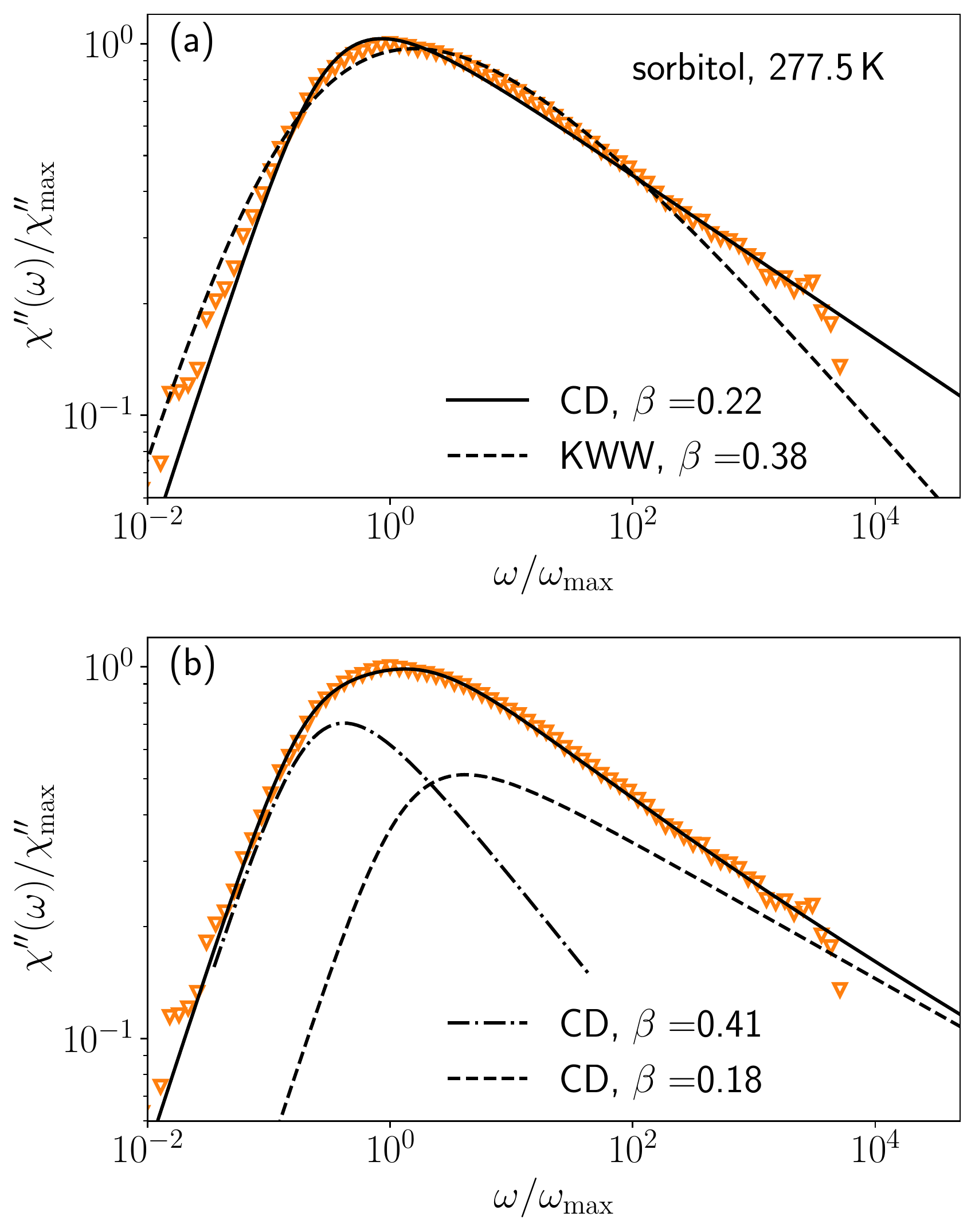}
    \caption{Different fits to the normalized sorbitol data at 277.5\,K from PCS. In (a) single model functions (CD and KWW) are used. Both fail to describe the slightly bimodal shape. The sum of two CD functions is used in (b) to qualitatively illustrate the bimodality.}
    \label{fig:sorbitol_shape}
\end{figure}

Interestingly, the spectrum of sorbitol displays signs of a bimodal $\alpha$-process peak structure. We attempt to illustrate this fact in Fig.\,\ref{fig:sorbitol_shape}, by fitting the normalized spectrum. In Fig.\,\ref{fig:sorbitol_shape}\,(a), we apply typical model functions, namely the Cole-Davidson (CD) and the Kohlrausch-Williams-Watts (KWW) function, which are routinely used to fit the $\alpha$-process near $T_\mathrm{g}$ in the literature. Both are unable to describe the entire spectrum, due to the shoulder on the high-frequency side of the peak. In Fig.\,\ref{fig:sorbitol_shape}\,(b), the sum of two CD functions is used, one to describe the low-frequency side of the spectrum that was shown to follow the universal shape in Fig.\,\ref{fig:pcs_shape}, and the second one to account for the additional high-frequency shoulder. This procedure is able to describe the entire spectrum. Of course, the better fit in (b) compared to (a) is not unexpected, simply considering the enhanced number of fit parameters used in (b). Therefore we do not want to base any conclusions on the fit procedure alone and for the moment the decomposition in Fig.\,\ref{fig:sorbitol_shape}\,(b) should merely be seen as an illustration.

\begin{figure}
    \centering
    \includegraphics[width=0.45\textwidth]{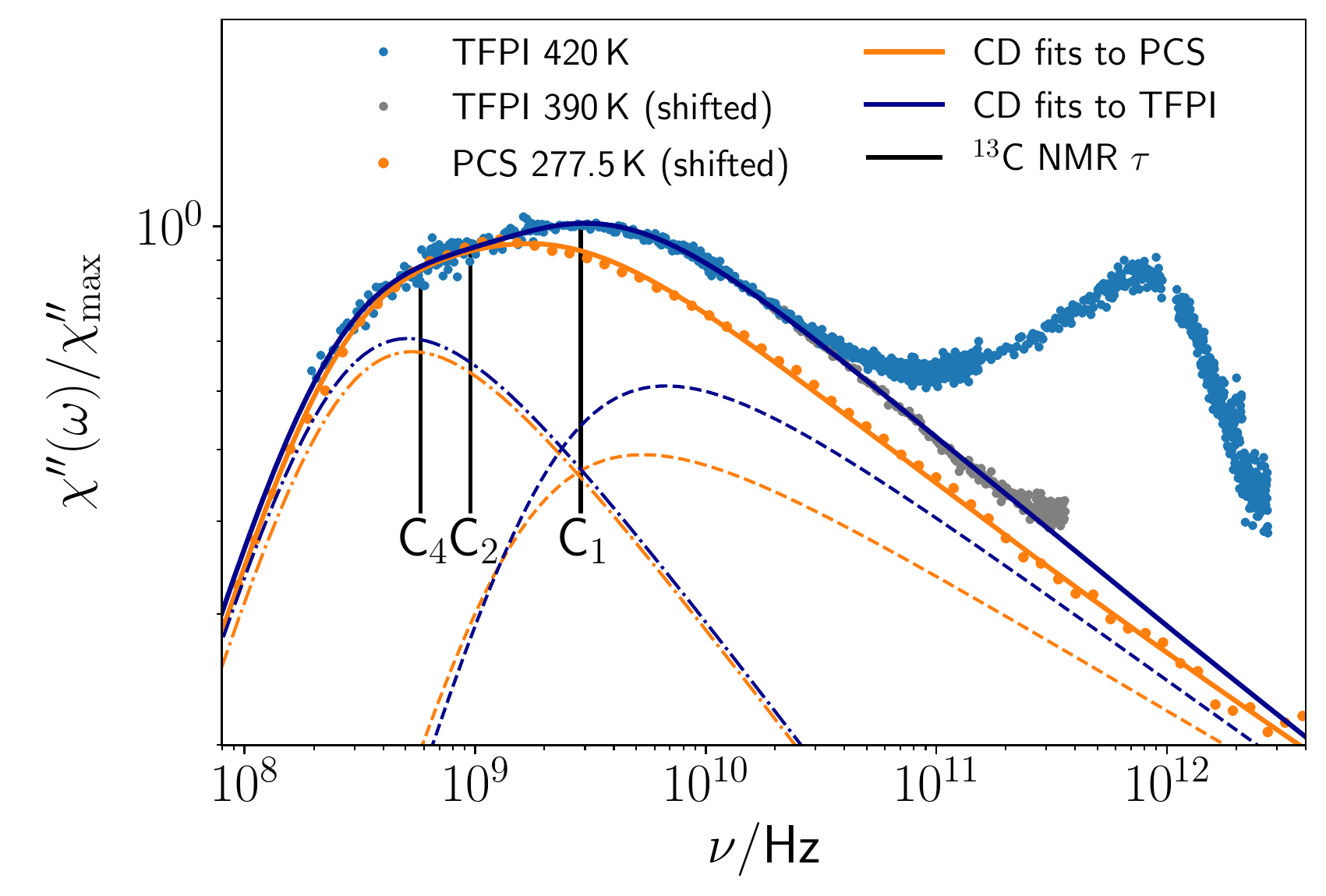}
    \caption{TFPI data of sorbitol at 420\,K (blue symbols) in the GHz region. TFPI data at 390\,K (gray symbols), shifted on the frequency axis, are added to illustrate the shape of the high-frequency power law, which is partly superimposed by the microscopic dynamics at 420\,K. PCS data at 277.5\,K (orange symbols) are added, shifted on the frequency axis. Fits of the sum of two CD functions to TFPI and PCS data are shown as blue and orange lines respectively. Dashed and dash-dotted lines each represent one of the two CD functions, whereas the solid lines represent the respective sums of both. Relaxation time constants from $^{13}$C-NMR experiments\,\supercite{Sixou2001} representing the dynamics of C-H bonds at the C$_1$, C$_2$ and C$_4$ carbon atom of sorbitol are indicated by the black lines.}
    \label{fig:sorbitol_tandem}
\end{figure}

One obvious explanation of a possible bimodal shape are intramolecular degrees of freedom, which are known to occur in sorbitol from NMR experiments and simulations, as already discussed above. Typically, dynamic decoupling of different intramolecular moieties in flexible molecules is pronounced in the liquid regime at high temperatures, and tends to be reduced in the supercooled regime, when cooperativity increases\,\supercite{Becher2021}. Therefore, we present TFPI data of sorbitol at high temperatures in Fig.\,\ref{fig:sorbitol_tandem}. At 420\,K (blue symbols), the $\alpha$-process is located in the GHz region and is thus, well resolved in the TFPI spectrum. In the THz region, microscopic dynamics is observed, which we do not want to focus on in this study. At high temperatures the $\alpha$-process is clearly bi- or even multimodal, indicated by the very broad peak, featuring two clearly distinct shoulders.

To find out whether this bimodality is related to intramolecular degrees of freedom, we include time constants from $^{13}$C-NMR  of the C$_4$-H, C$_2$-H and C$_1$-H bonds (numbered along the carbon backbone) from Ref.\,\cite{Sixou2001} as black lines. To obtain time constants at exactly 420\,K, we extrapolate these data, utilizing the fact that their temperature dependence is approximately Arrhenius in the GHz region (see SI). Considering that time constants of different molecular sites are expected to be identical in case of isotropically rotating rigid molecules, the deviations indicate that the sorbitol molecule is markedly flexible and different molecular moieties rotate on different timescales. Interestingly, the NMR time constants coincide with the positions of the two shoulders that are observed in the TFPI spectrum. This agreement clearly suggests that, indeed, the dynamic decoupling of different molecular moieties leads to the strong broadening and the bimodality of the TFPI spectrum. This is not surprising, as rotation of the outer parts (C$_1$ and C$_6$) of the sorbitol molecule will have a significant impact on the optical anisotropy tensor, as they make up a third of the molecule. Additionally, the dynamics of O-H groups is markedly heterogeneous\,\supercite{Becher2021}, decouples from the dynamics of C-H bonds\,\supercite{Doess2001} and thus, can also contribute to the TFPI spectrum. Especially the broadening at $\nu>\nu_{\mathrm{C1}}$ could result from O-H dynamics, suggested by the very asymmetrically broadened spectrum found in $^2$H-NMR for OD-labeled sorbitol\,\supercite{Becher2021}.

To compare the high- to the low-temperature dynamics, we include a representative PCS spectrum in Fig.\,\ref{fig:sorbitol_tandem} as the orange symbols. It is shifted along the frequency axis so that the low-frequency part of TFPI and PCS spectrum approximately coincide. Both spectra are comparably broad and share a very similar high-frequency power law. To emphasize the latter, we add a second TFPI spectrum, measured at 390\,K as the grey symbols, shifted along the frequency axis to extend the frequency range the power law is observed in. In addition, we fit the TFPI spectrum using two CD functions (blue lines) and compare them to the fits to the PCS spectrum (orange lines) from Fig.\,\ref{fig:sorbitol_shape}(b). Both spectra are well-described by similar sets of parameters for the two CD functions. The difference between PCS and TFPI spectrum primarily seems to be the intensity of the faster contribution, which is larger for the high temperature measurement. From the comparison of high- and low-temperature DLS data we therefore conclude that intramolecular dynamic heterogeneity has a major impact on the spectral shape of sorbitol.
Although the impact of intramolecular degrees of freedom is smaller at low temperatures, the sorbitol molecule seems to be flexible enough so that dynamic decoupling of different molecular moieties to some degree persists even in the cooperative regime close to $T_\mathrm{g}$.

\begin{figure*}[h!]
\centering
  \includegraphics[width=0.9\textwidth]{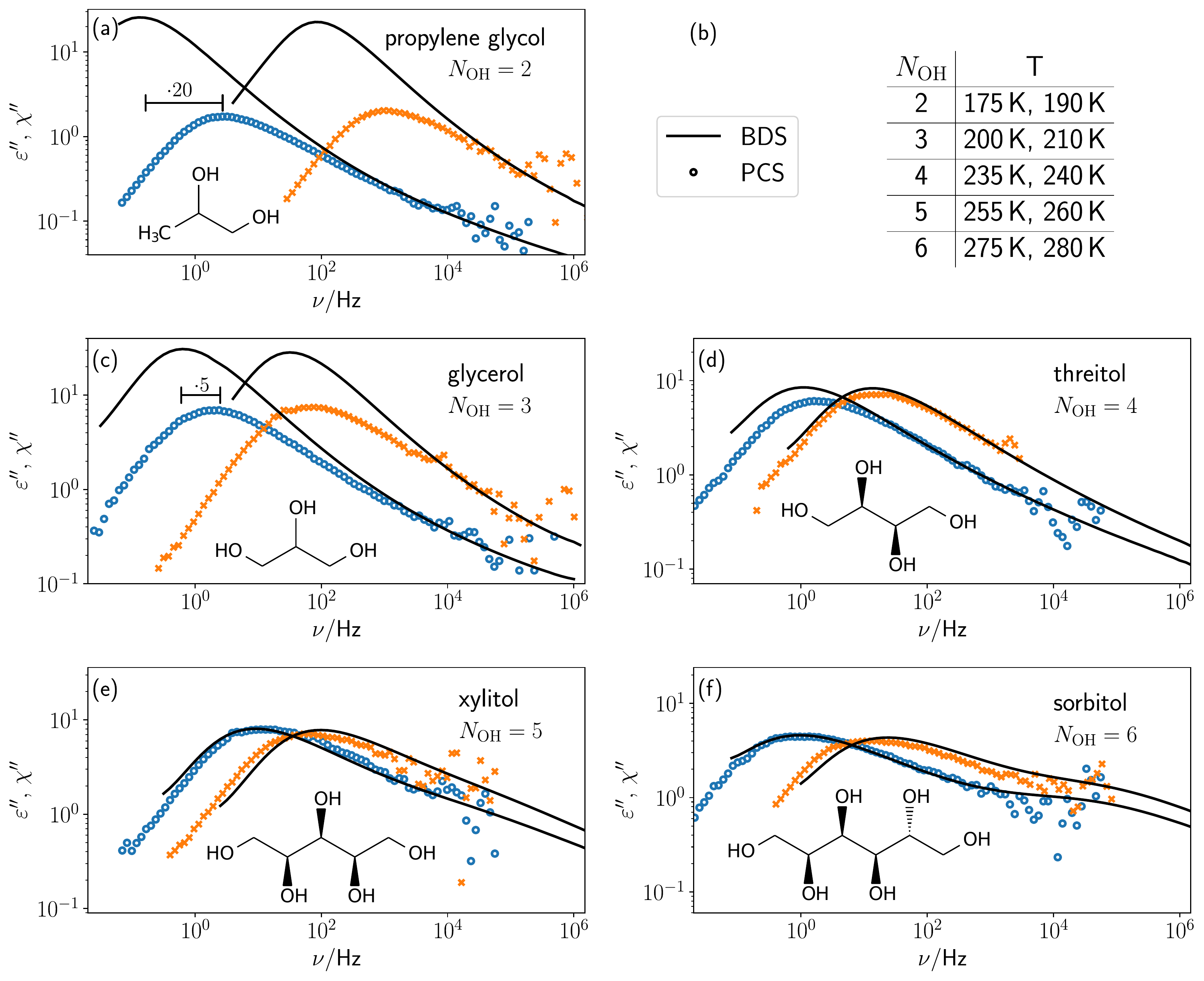}
  \caption{Comparison of BDS (lines) and PCS data (symbols) at the selected temperatures shown in (b). For PG and glycerol the factor between the peak maximum frequencies from BDS and DLS are indicated.}
  \label{fig:comp_bds_pcs}
\end{figure*}

As a next step, in Fig.\,\ref{fig:comp_bds_pcs} we now compare PCS and BDS spectra at the same temperatures. For every PA we include two temperatures (see table in Fig.\,\ref{fig:comp_bds_pcs}(b)) to exclude the possibility of any significant difference in temperature dependence of BDS and PCS data. As the absolute amplitude of $\chi^{\prime\prime}$ can not easily be determined, we adjust it in order for the high-frequency region of BDS and DLS data to coincide. We note that for glycerol the absolute intensity of $\chi^{\prime\prime}$ was determined in an elaborate analysis, finally confirming the explained procedure of adjusting the amplitude for this particular liquid\,\supercite{Gabriel2020}. Starting with PG and glycerol, two observations can be readily made: First, the peak maxima of $\chi^{\prime\prime}$ are located at higher frequencies compared to those of $\varepsilon^{\prime\prime}$. Both are separated by a factor 20 for PG and a factor 5 in the case of glycerol. Second, $\beta_\mathrm{BDS}>\beta_\mathrm{DLS}=0.5$ holds in both cases, with $\beta_\mathrm{BDS}\approx 0.7$ for PG and $\beta_\mathrm{BDS}\approx 0.6$ for glycerol. Both observations are in line with the assumption that slow dipolar cross-correlations contribute to the dielectric loss spectra as a Debye-like process. An elaborate analysis and curve-fits are shown in Ref.\,\cite{Gabriel2020} for glycerol and can be found in the SI of the present paper for PG. For threitol, BDS and PCS data show almost the same relaxation times with $\tau_\mathrm{BDS}$ being slightly longer than $\tau_\mathrm{PCS}$, the difference being hardly significant considering the temperature uncertainty of $\pm0.5\,$K between both setups. Also, $\beta_\mathrm{BDS}$ approaches $\beta_\mathrm{PCS}$. For xylitol and sorbitol, BDS and PCS data coincide regarding peak maximum frequency and also roughly regarding high-frequency power law. However, in both cases the PCS spectra display signs of bimodality, which are not clearly resolved in BDS. The slight shifts between BDS and PCS spectra of xylitol and sorbitol at the respective higher temperature most likely originate from  small temperature deviations combined with the large fragility\,\supercite{Doess2002} of these two PAs.

\section{Discussion}
To discuss the presented results, we want to focus on the three questions raised in the introduction of this paper.
\paragraph*{Generic shape of the $\alpha$-relaxation in PAs}
When analyzing the PCS data, only short-chained PAs display the generic shape of the $\alpha$-relaxation proposed in Ref.\,\cite{Pabst2021}. By contrast, the longer-chained PAs follow the generic shape only up to frequencies slightly above the peak maximum and deviations are observed at higher frequencies that systematically increase with increasing chain-length. 
For sorbitol these deviations were shown to be caused by intramolecular dynamic heterogeneity. The latter is even more pronounced at higher temperatures in the liquid regime and can, e.g., be attributed to the decoupled mobility of C-H bonds at different positions in the sorbitol molecule\,\supercite{Sixou2001}. Furthermore, it is known that also O-H bonds dynamically decouple from other bonds in the molecule\,\supercite{Doess2001,Becher2021}. Essentially, the dynamic decoupling of different bonds reflects the fact that conformational transitions within the molecule take place on similar timescales as the $\alpha$-process. By contrast, in short-chained PAs, namely glycerol and PG, relaxation times of different bonds are known to coincide and simulations reveal that conformational transitions are slow compared to the $\alpha$-relaxation\,\supercite{Becher2021a}. Consequently, short-chained PAs rotate as a rigid entity. Although no respective NMR or simulation data exist for threitol and xylitol, it is straight forward to assume that some transition occurs between the short- and the long-chain behavior. This implies that somewhere between $N_\mathrm{OH}=3$ and $N_\mathrm{OH}=6$, the characteristic timescale of conformational transitions approaches the one of the $\alpha$-process, thus leading to deviations from the generic shape. Since the DLS spectrum of threitol slightly deviates from the generic shape, intramolecular dynamic heterogeneity in PAs seems to become relevant around $N_\mathrm{OH}=4$.

At this point is is worth noting that the vast majority of molecules, whose DLS spectra were shown to follow the generic shape\,\supercite{Pabst2021}, are in fact to some degree flexible. However, usually intramolecular degrees of freedom are either very fast compared to the $\alpha$-process and, e.g., appear as a non-Johari-Goldstein $\beta$-process\,\supercite{Kaminska2007}, or are frozen in the deeply supercooled and cooperative regime\,\supercite{Becher2021}. Therefore, it is necessary to address the question why conformational transitions and the $\alpha$-process appear on comparable timescales in longer-chained PAs. From our perspective, two aspects are important to consider in this regard:

First, the potential energy landscape of sorbitol is quite complex, due to the existence of multiple interaction mechanisms and, most importantly, due to the large number of intra- and intermolecular hydrogen bonds involved\,\supercite{Ruggiero2017,Krynski2020}. The complexity gives rise to several conformational states of sorbitol separated by, presumably, low potential energy barriers, which finally favours transitions between these conformations.

Second, due to the number of hydrogen bonds a sorbitol molecule is involved in, a reorientation of the carbon backbone requires several hydrogen bonds at different positions of the molecule to break simultaneously, which is quite unlikely. Therefore, smaller segments of the molecule, e.g. one or two C-O-H segments, are more likely to rotate independently, thus the $\alpha$-process in fact occurs via a series of conformational transitions. In such a scenario, bonds at the end of the molecule rotate slightly faster than the ones in the center and O-H bond dynamics can decouple from C-H or C-C bond dynamics, all of which was observed in the case of sorbitol.\,\supercite{Sixou2001,Doess2001,Becher2021}

We therefore conclude that only under special circumstances intramolecular dynamic heterogeneity can significantly affect the shape of the $\alpha$-process. These conditions seem to be fulfilled in the case of longer-chained PAs, therefore the additional intramolecular dynamic heterogeneity leads to a broader spectral shape compared to the generic behavior.

Finally, it remains to be discussed to what extent intramolecular dynamic heterogeneity appears in the \textit{dielectric} loss spectra of xylitol and sorbitol. In BDS, only the rotation of C-O-H groups is resolved, due to the position of the dipole moment, and rotations of other bonds are only probed indirectly in case they also induce the motion of a C-O-H bond. Therefore, DLS and BDS spectra do not necessarily reflect intramolecular dynamic heterogeneity in an equal way and this, indeed, seems to be the case, as the bimodality, observed for the DLS spectra, cannot be identified directly for the BDS spectra. Nevertheless, as both spectra are still very similar, it seems likely that the very broad shape of the dielectric loss peak in sorbitol at least to some degree results from intramolecular dynamic heterogeneity of different C-O-H bonds.

\paragraph*{Slow cross-correlations in the dielectric spectra of PAs}
As discussed above, slow dipolar cross-correlations which lead to a Debye like process have been observed in the dielectric spectra of both, H-bonding and non-H-bonding polar liquids\,\supercite{Pabst2021}, in line with a recent theory of electric polarization\,\supercite{Dejarding2019}. The same effect is observed for PG and glycerol, considering the differences between their respective BDS and PCS spectra. We observe that the slow cross-correlations are stronger and more separated in PG than in glycerol. For threitol small cross-correlation effects might contribute, while for xylitol and sorbitol no contributions slower than the PCS spectra are observed. Thus, the strength of slow cross-correlations decreases with increasing  chain-length of PAs.

This effect can be explained, considering that intramolecular connectivity between dipolar subunits will to some degree hinder the formation and longevity of dipolar intermolecular cross-correlations, so that a Debye-like process is only detected if the dipole moment is to a large extent fixed to the molecular frame. For PG and glycerol, the dipole moment is fixed to the molecule, as the molecules rotate approximately as a rigid entity\,\supercite{Becher2021a}. Therefore, in these liquids dipolar cross-correlations can persist up to timescales longer than the $\alpha$-process, in line with the experimental results. For xylitol and sorbitol it was shown that conformational transitions occur on the timescale of the $\alpha$-process. Due to the fact that the direction and the strength of the molecular dipole moment strongly depends on the conformational state of the molecule\,\supercite{Alonso2018}, dipolar cross-correlations will relax on similar timescales as the $\alpha$-process. In this case, the contribution of cross-correlations to the dielectric spectra can not be straight-forwardly identified from comparison to DLS spectra, as they mainly modify the amplitude of the dielectric $\alpha$-process. However, they can still be identified in the static permittivity, considering that an attempt to quantify the Kirkwood correlation factor $g_\mathrm{k}$ of sorbitol yielded values $>1$\,\supercite{Nakanishi2010}, albeit an exact calculation of $g_\mathrm{k}$ is difficult as the total molecular dipole moment is ill defined due to intramolecular flexibility. We therefore conclude that while static dipolar cross-correlations are found in sorbitol, and presumably also xylitol, they are not particularly long-lived and do not lead to a separate slow process, thus are not identified by comparing BDS and DLS spectra.

\paragraph*{Conditions for observing the generic shape}
Finally, we want to address the conditions for observing the generic shape of the $\alpha$-process in supercooled liquids. Typically, the generic shape is only observed in the deeply supercooled liquid (mHz-kHz), but not in the weakly supercooled regime (GHz-THz)\,\supercite{Schmidtke2014,Pabst2021}. It is therefore straight forward to assume that the appearance of the generic shape is in some way related to glassy cooperativity, which develops in the deeply supercooled regime, possibly below a certain characteristic temperature as several theories and models indicate \,\supercite{Cavagna2009}. This is consistent with the observation that the high-frequency dependence in dielectric spectra of different supercooled liquids converges to the generic $\omega^{-1/2}$ power law with decreasing temperature\,\supercite{Nielsen2009} and thus, with increasing cooperativity. 

Of course, glassy cooperativity is first of all an \textit{inter}molecular phenomenon. Accordingly, theories predicting the $\omega^{-1/2}$ high-frequency power law, do not deal with any intramolecular peculiarities\,\supercite{Dyre2005,Dyre2005a}. Also, from an experimental point of view neither intramolecular properties nor particular intermolecular interactions seem to play a role, considering the broad diversity of supercooled liquids that were all confirmed to follow the generic shape. However, it seems to be important in order to observe the generic shape that the experimental technique of choice actually probes only the \textit{inter}molecular distribution of relaxation times and no aspects of \textit{intra}molecular dynamics. Seemingly, in many instances this is the case for DLS. First, because the technique in general probes the reorientation dynamics averaged over the entire molecule, since all molecular units contribute to the polarizability tensor. There may be exceptions where highly anisotropic groups, like, e.g., phenyl rings are involved, which may dominate the scattering signal in some cases. Second, because DLS seems to be mostly insensitive to slow dipolar cross-correlation contributions that can mask the generic shape of the $\alpha$-relaxation. BDS, by contrast, is very sensitive to slow dipolar cross-correlations\,\supercite{Pabst2020,Pabst2021}, and is, at the same time, insensitive to apolar groups, which can make up a large portion of the molecule, especially in organic systems. $^2$H-NMR, to name another technique,  is truly insensitive to cross-correlations. However it explicitly probes specific chemical bonds in the molecules and, thus, probes molecular dynamics in a very local fashion. This can result in probing different spectral shapes compared to DLS, and to observe the generic shape requires the molecule to be actually rigid and the reorientation to be isotropic.

Considering the discussed aspects, one can only expect to observe the generic spectral shape in cases where intramolecular dynamic heterogeneity does not play a role on timescales comparable to the $\alpha$-relaxation. These intramolecular degrees of freedom seem to be the reason why long-chained PAs deviate from the generic shape. However, these effects only appear under special circumstances, as intramolecular degrees of freedom usually are frozen in the highly cooperative regime close to $T_\mathrm{g}$. 

\section{Conclusions}
Using BDS and DLS, we have studied a systematic series of PAs, referred to in the literature as prototypical glass-formers. The main objectives have been to verify, whether the recently proposed generic shape of the $\alpha$-process\,\supercite{Pabst2021} is also observed in PAs, and whether slow dipolar cross-correlations contribute to the dielectric spectra of all PAs in a similar way as previously shown for glycerol\,\supercite{Gabriel2020}. In these regards, PAs separate into two categories: On the one hand, short-chained PAs, namely PG and glycerol, follow the generic shape and display significant contributions of slow cross-correlations to their respective BDS spectra. On the other hand long-chained PAs, namely xylitol and sorbitol, deviate from the generic shape and show no signs of slow cross-correlations. Threitol, the intermediate chain-length PA, takes a place in between both categories, as slight deviations from the generic shape and a possible small contribution of slow cross-correlations are observed. 

The situation in long-chained PAs can be understood by considering intramolecular dynamic heterogeneity, as we have shown by comparing low- and high-temperature DLS spectra and time constants from $^{13}$C-NMR of different dynamically decoupled C-H bonds. These effects enhance the overall dynamic heterogeneity, thus leading to broader relaxation spectra and to the relaxation of dipolar cross-correlations on timescales comparable to the one of the $\alpha$-process. In contrast, short-chained PAs behave rigid on the relevant timescales, and consequently generic behavior and a separable Debye-like cross-correlation peak are observed.

Finally, we attempted to identify possible conditions for observing a generic shape in DLS and explain why deviations from this shape can be observed in other experimental techniques. To observe the generic shape, the probed dynamics need to reflect the reorientation of a molecule as a whole and thus the intermolecular distribution of relaxation times. This is not the case if slow cross-correlations or intramolecular motions superimpose with the $\alpha$-process and deviations might also occur if the experimental technique of choice selectively probes the dynamics of specific parts of a molecule.

\section*{Author Contributions}
Th.B., J.G. and T.B. conceptualized and supervised experiments, T.B., J.G, T.R. and R.Z performed experiments and analyzed data, T.B. wrote the original draft, Th.B., J.G, T.R. and R.Z. edited and reviewed the draft.

\section*{Conflicts of interest}
There are no conflicts to declare.

\section*{Acknowledgements}
The authors thank Florian Pabst for carefully reading the manuscript and useful remarks. Financial support by the Deutsche Forschungsgemeinschaft
under grant no. BL 1192/3 and by the VILLUM~Foundation’s \emph{Matter} grant under grant~no.~16515 is gratefully acknowledged.

\printbibliography

\end{document}